# Beyond Visibility and Technical Reuse: Public Application Transformation in Open-Source Model Ecosystems

Duorong Wang[1,2], Xiaoting Wei[1], Bin Liu[1], Jiannan Yang[1,*]

1. School of Information Management, Nanjing University

2. Division of Emerging Interdisciplinary Areas, The Hong Kong University of Science and Technology

*Corresponding author: Jiannan Yang (jnyang@nju.edu.cn)

## Abstract

Open-source model platforms have made it easier to publish AI models, but model release alone does not reveal whether models become visible, technically reused, or incorporated into public applications. This study introduces public application transformation as a platform-visible dimension of model impact and examines it through structured Model–Space links on Hugging Face. We construct a platform-scale dataset of 2.56 million model repositories, 1.06 million Spaces, 810,087 dataset repositories, and 1.22 million account profiles, together with Model–Space, Dataset–Space, and model-to-model technical reuse links. The analysis shows that public application transformation is highly selective and concentrated: only a small share of models are linked to Spaces, and most Model–Space links are concentrated among a limited set of models. More importantly, application transformation is associated with platform visibility but is not equivalent to technical reuse, indicating that downloads, likes, downstream model reuse, and application-facing uptake capture different forms of model impact. Additional analyses show that application-transformed models tend to exhibit stronger metadata-based readiness and enter heterogeneous Space configurations involving datasets, SDKs, and task-specific application categories. By tracing how models move from repositories into public applications and demos, this study extends the measurement of open-source model impact from artifact availability and technical reuse to platform-mediated transformation across AI information objects.

## Introduction

Open-source model platforms have made it easier than ever to publish AI models, but the abundance of released models has also made model impact harder to interpret. On platforms such as Hugging Face, models can be released, updated, downloaded, liked, fine-tuned, merged, quantized, and connected to other platform resources at large scale. This expansion has made open-source model ecosystems increasingly important as infrastructures for AI development and reuse (Shi & Jiang, 2025). At the same time, openness itself cannot be reduced to the mere availability of model files, since the practical value of open-source AI also depends on whether models can be inspected, reused, extended, and embedded in downstream systems (Rehak et al., 2025). As a result, model release is no longer a strong signal that a model has been widely noticed, reused, or incorporated into applications. Many models remain isolated repositories after publication, while a smaller number become upstream inputs for downstream model development or components of public-facing applications and demos. The growth of open-source model ecosystems therefore raises a measurement problem: how can we distinguish models that are merely available from models that travel across platform layers and become useful in broader technical or application contexts?

Existing indicators capture only parts of this trajectory. Downloads and likes measure platform attention (Shi & Jiang, 2025), but they do not show how a model is used. Benchmark scores measure performance under standardized evaluation settings (P. Liang et al., 2022; Srivastava et al., 2023), but they do not show whether a model is adopted in downstream contexts. Model lineage, base-model references, and derivative repositories capture technical reuse, but they mainly describe how models are reused to produce other models (Taraghi et al., 2024). Prior work on platform metrics has similarly shown that popularity, visibility, adoption, and reuse indicators often capture different forms of platform activity rather than serving as equivalent proxies for the same kind of impact (Saini et al., 2020; Xie et al., 2021; Zerouali et al., 2019). Recent work on open-source foundation model ecosystems and Hugging Face model reuse has similarly emphasized downloads, derivative innovation, downstream repositories, and model reuse practices as important signals of model impact (Jiang et al., 2024; Shi & Jiang, 2025; Taraghi et al., 2024). These indicators are important, yet they leave a gap between model release and application-facing use. A model may attract many downloads without appearing in public applications. Another model may be technically reused by many downstream models but remain largely invisible in application interfaces. Conversely, some models may become frequently incorporated into demos even if they are not the most prominent upstream models. These differences suggest that open-source model impact is not a single hierarchy, but a set of distinct pathways.

One under-measured pathway is the movement from model repositories into public applications and demos. This transition is not the same as commercial deployment or production adoption. It is a more specific platform-visible event in which a model becomes part of an application-facing artifact, such as an interactive demo, interface, benchmark app, tool, or resource configuration. On Hugging Face, this pathway is partly observable through Spaces. Prior quantitative work on Hugging Face has already shown that the Hub contains not only models and datasets, but also Spaces and related activity patterns (Osborne et al., 2024). System and demo papers also illustrate how Hugging Face Spaces can host public interfaces and tools around model or data resources (Arora et al., 2025; Piktus et al., 2023). A Model–Space link therefore indicates that a model has crossed an observable boundary: it is no longer only a repository object, but has been incorporated into a public Space where it can be demonstrated, combined with other resources, or made available through an interface.

Against this background, this study examines public application transformation in the Hugging Face ecosystem. We focus on three research questions. First, how is public application transformation structured and temporally organized in an open-source model ecosystem? Second, how does public application transformation relate to, and differ from, community visibility and technical reuse? Third, what application-readiness signals and resource assemblages are associated with this transformation after models enter the Space layer? To answer these questions, we construct a platform-scale dataset of Hugging Face models, datasets, Spaces, accounts, and structured relations. The dataset includes approximately 2.56 million model repositories, 1.06 million Spaces, 810,087 dataset repositories, and 1.22 million account profiles. We derive Model–Space links, Dataset–Space links, and model-to-model technical reuse relations, and use these relations to compare application transformation with downloads, likes, technical reuse, metadata readiness, and Space-level resource configurations.

The analysis shows that public application transformation is sparse, concentrated, visibility-dependent, and not reuse-equivalent. Only a small minority of models are linked to Spaces, and Model–Space links are highly concentrated among a limited subset of models. At the same time, these links are not simply another expression of downloads, likes, or model-to-model reuse. Public application transformation overlaps with conventional impact indicators, but it captures an application-facing profile of model impact that remains distinct from technical reuse. Models are more likely to enter Spaces when they provide clearer task, library, and license metadata, extending prior concerns that model documentation and metadata shape how models can be understood, compared, and reused. After entering Spaces, models are organized into heterogeneous resource assemblages involving datasets, SDKs, author relationships, and task-specific application categories. By measuring this pathway, the study provides a way to evaluate open-source model impact beyond release counts, platform attention, and technical reuse, and to understand how models become transformed into public application-facing artifacts.

## Literature Review

### Open-source model ecosystems as platform infrastructures

Open-source software research has long shown that repositories and hosting platforms do more than store artifacts (Buckland, 1991). They organize distributed production, collaboration, visibility, and reuse through contributors, maintainers, licenses, documentation, dependencies, and platform signals (Crowston et al., 2008; Petralia, 2025). Open-source model ecosystems extend this platform logic while changing the nature of the shared artifact. Models are not only code repositories; they are accompanied by weights, configuration files, model cards, datasets, task labels, libraries, licenses, evaluation records, and sometimes application interfaces. As a result, repository release alone provides only a partial view of model circulation. A released model becomes reusable only when surrounding platform resources make it discoverable, interpretable, executable, and governable (Candela et al., 2015).

This layered organization has made openness in AI more complex than simple public availability. Studies of generative AI release practices describe openness as a gradient rather than a binary condition (Solaiman, 2023), while related work argues that meaningful open-source AI depends on the ability to inspect, adapt, extend, and evaluate models in practice (Rehak et al., 2025). Hugging Face is a prominent example of this platformed model ecosystem. The Hub brings together models, datasets, and Spaces within a shared environment, and recent studies have documented its rapid growth, diverse participation, industrial relevance, and documentation challenges (Osborne et al., 2024; Sangari et al., 2025). These studies establish model hubs as important infrastructures for open-source AI development. They also point to a further challenge: models circulate through multiple platform layers, including visibility metrics, technical reuse relations, documentation practices, and application-facing interfaces. How these layers differ, and how models move among them after release, remains less clearly specified.

### Measuring model impact: visibility, technical reuse, and application transformation

Impact measurement on digital platforms often begins with traces of attention, use, and reuse, such as views, downloads, citations, stars, forks, dependencies, and user activity (Silvello, 2018; Sugimoto et al., 2017). In software and data repository studies, these indicators have been used to measure visibility, popularity, adoption, and reuse because they make large-scale platform activity comparable across artifacts (Xie et al., 2021). However, prior work also shows that different platform indicators capture different forms of activity and should not be treated as equivalent proxies for the same kind of impact. Studies of software package ecosystems, for example, find that artifact rankings vary depending on whether impact is measured through downloads, dependencies, GitHub activity, or other popularity metrics (Borges et al., 2016; Saini et al., 2020; Zerouali et al., 2019). Related work on data repositories similarly shows that metadata, repository context, downloads, stars, forks, and citations reflect different pathways through which shared artifacts become visible and valued (Xie et al., 2021). This literature provides an important baseline for studying open-source model ecosystems: platform indicators should be interpreted as traces of specific forms of activity, not as interchangeable measures of a single underlying impact hierarchy.

The same measurement problem becomes more complex when the shared artifacts are machine learning models. Models can attract downloads and likes, achieve benchmark visibility, serve as upstream inputs for downstream model development, or be incorporated into tools and applications. Recent work has begun to develop impact evaluation frameworks for open-source foundation model ecosystems, emphasizing visibility, user engagement, derivative innovation, and ecosystem participation as distinct dimensions of model influence (Shi & Jiang, 2025). A particularly important dimension is technical reuse. In model hubs, existing models are frequently fine-tuned, adapted, merged, quantized, or otherwise reused as upstream inputs for new model repositories. Studies of Hugging Face model reuse show that these practices are central to the platform's development dynamics and that developers face both benefits and challenges when reusing deep learning models (Taraghi et al., 2024). Other datasets and empirical studies extend the analysis beyond model-hub metadata by linking pre-trained models to downstream open-source software repositories, thereby tracing how models appear in code-level use and developer practices (Jiang et al., 2024; Palombo et al., 2025). Together, these studies establish technical reuse as a major form of model impact, especially when models become inputs for further model production or software development.

However, technical reuse does not exhaust downstream transformation. A model may be reused to produce another model without becoming visible through an application interface, while another model may appear frequently in public demos without becoming a major upstream source for derivative model development. This distinction is reinforced by work on model discovery and repository organization, which shows that reuse depends on the ability to search, compare, and evaluate models through metadata, descriptions, and task information (Li, Qi, et al., 2023; Zhang et al., 2024). It is also visible in the growing use of Hugging Face Spaces, which host public demos, tools, benchmark interfaces, and interactive applications that can explicitly reference models and datasets. Prior quantitative work on Hugging Face Hub recognizes Spaces as part of the platform's activity structure, while system and demo papers show how Spaces can be used to host public model- or data-oriented interfaces (Osborne et al., 2024; Piktus et al., 2023). What remains less clearly specified is how the movement from model repositories into these public application-facing objects differs from visibility or model-to-model technical reuse at ecosystem scale. This gap motivates treating visibility, technical reuse, and application transformation as related but analytically distinct pathways of open-source model impact.

**Documentation, metadata, and application readiness**

As model repositories multiply, the usability of a model increasingly depends on the information that surrounds it. A released model is difficult to evaluate or reuse from weights alone. Developers need to know what the model is intended to do, how it can be loaded or integrated, what limitations it has, and under what conditions it can be reused. Work on model cards established documentation as a mechanism for reporting model characteristics, intended uses, limitations, and ethical considerations (Gebru et al., 2021; Mitchell et al., 2019). Subsequent studies show that documentation practices remain uneven across AI repositories. Large-scale analyses of Hugging Face model cards find substantial variation in what information is reported, while studies of intended usage context show that many model repositories provide incomplete or ambiguous information about how models should be used (Gong et al., 2023; W. Liang et al., 2024). These findings indicate that documentation is not only a transparency issue. It also shapes whether models can be understood and reused by actors beyond their original developers.

Structured metadata provides a related form of readiness. Task labels help users identify what a model is designed for. Library metadata connects a model to software environments and inference routines. License metadata clarifies reuse conditions and reduces uncertainty for downstream developers. Research on queryable model repositories has emphasized that structured metadata is essential for model search, comparison, and retrieval (Li, Kant, et al., 2023). Work on AI bill-of-materials, ethical documentation, and repository transparency further shows that metadata and documentation support governance, accountability, and supply-chain understanding in open-source AI systems (Gao et al., 2024; Silvello, 2018; Stalnaker et al., 2025; Wilkinson et al., 2016). These studies suggest that metadata is not merely descriptive. It helps make models findable, interpretable, integrable, and governable within platform environments.

This literature provides the basis for thinking about application readiness. Existing work has primarily examined documentation quality, metadata completeness, retrieval, and transparency as repository-level concerns. Less is known about whether these informational signals are associated with movement into application-facing contexts. A model that lacks task metadata, license information, or library compatibility may still be technically capable, but it may be harder to discover, evaluate, and package into a public demo or tool. Conversely, models with clearer metadata may be better positioned to travel beyond repository-level availability. The question is therefore not only whether open-source models are documented, but whether documentation and metadata are associated with their transformation into public applications and demos.

## Methods

### Data construction

This study uses publicly available metadata from Hugging Face to construct a platform-scale dataset of open-source model repositories, dataset repositories, Spaces, and account profiles. We collected repository metadata through the platform's public interfaces and structured metadata fields. As of February 3, 2026, the dataset contains 2,556,240 model repositories, 1,063,489 Spaces, 810,087 dataset repositories, and 1,219,857 account profiles. These records provide a large-scale snapshot of the Hugging Face ecosystem and allow us to examine how models, datasets, and public application-facing Spaces are connected through explicit metadata relations.

Following Hugging Face's own organization of the Hub, we treat models, datasets, and Spaces as three primary repository-level objects. Model repositories are used to host and share models across LLM, text, vision, audio, and multimodal tasks. Their metadata include repository identifiers, authorship, timestamps, engagement indicators, task labels, library information, license information, and selected model-card or configuration fields. Dataset repositories provide data resources across domains, tasks, languages, and modalities, with metadata describing authorship, downloads, tags, licenses, language, task, format, and size-related information. Spaces are interactive applications or demo apps for presenting machine learning models in the browser. Their metadata include authorship, timestamps, Software development kit (SDK) information, tags, likes, and structured references to models and datasets. Account profile metadata were also collected and linked to repository authors where available, allowing us to distinguish individual and organizational accounts.

The repository-level records were converted into normalized entity and relation tables. The canonical entity tables contain one row per model, dataset, Space, or account. From structured metadata fields, we extracted four main types of links. A Model–Space link is recorded when a Space explicitly lists a model in its structured *models[]* field. This relation serves as the main empirical trace of public application transformation. A Dataset–Space link is recorded when a Space explicitly lists a dataset in its structured *datasets[]* field, indicating whether a Space combines models with data resources. A model-to-model technical reuse link is recorded when one model identifies another model as its upstream technical input, such as a base model for fine-tuning, a source model used in merging, an adapter-trained variant, or a quantized derivative. A model–dataset link is recorded when model metadata explicitly references a dataset. In the processed dataset, we extracted 667,817 structured Model–Space links, 54,681 structured Dataset–Space links, 741,419 model-to-model technical reuse links, and 443,040 model–dataset links.

All entity and relation tables were processed using consistent normalization rules. Repository identifiers were standardized across models, datasets, Spaces, and accounts. Empty strings, placeholder values, and null-like labels were removed from relation records. JSON-like fields were parsed where necessary, and relation tables were deduplicated at the appropriate entity-pair level. We also normalized model-level task, library, and license fields into separate tables for constructing application-readiness measures. These processed tables form the empirical basis for analyzing public application transformation, community visibility, technical reuse, application readiness, and resource assemblages in the Hugging Face ecosystem.

**Measures and analytical variables**

Based on the normalized entity and relation tables, we constructed analytical variables at the model, Space, and Model–Space edge levels. These variables measure public application transformation, community visibility, technical reuse, application readiness, and resource assemblage. Table 1 summarizes the main constructs, operational measures, and units of analysis.

**Table 1. Key constructs and operational measures.**

| Construct | Operationalization | Unit of analysis |
|---|---|---|
| Public application transformation | Whether a model is linked to at least one Space; number of linked Spaces; number of cross-author Space links | Model; Model–Space edge |
| Transformation timing | Difference between model creation date and Space creation date | Model–Space edge |
| Community visibility | Downloads, likes, and a combined visibility score based on percentile ranks | Model |
| Technical reuse | Number of downstream model repositories that reuse the focal model as an upstream technical input | Model |
| Application readiness | Availability of task, library, and license metadata; readiness score; task family; reported size group; author type | Model |
| Application pathway | Task-to-application-category flow; same-author and cross-author use; dataset co-use by application category | Model–Space edge; Space |

Public application transformation is the main outcome of interest. A model is coded as application-transformed if it has at least one structured Model–Space link. We use a binary indicator, *has_application_transformation*, to measure whether a model enters this public application pathway, and *n_spaces* to measure the number of Spaces linked to the model. We also identify a Model–Space link as cross-author when the author account of the Space differs from the author account of the linked model repository. These links exclude cases in which model publishers create their own Spaces to showcase their models, and therefore provide a stricter trace of application-oriented uptake by other platform actors. For temporal analyses, we calculate the lag between model creation date and Space creation date. Because the exact time when a model reference was added to a Space is not observed, this lag measures relative timing between model release and Space creation rather than exact link-formation time.

Community visibility is measured through downloads, likes, and a combined visibility score. The combined score is defined as the maximum of a model's percentile rank in downloads and its percentile rank in likes, so that a model is treated as highly visible if it ranks highly on either dimension of platform attention. Technical reuse is measured by the number of downstream model repositories that reuse the focal model as an upstream technical input. This measure captures model-to-model reuse in the production layer, while Model–Space links capture application-layer reuse. Application readiness captures whether a model provides metadata that may facilitate discovery, interpretation, integration, and reuse in public applications or demos. We operationalize readiness using three binary indicators: whether a model has task metadata, library metadata, and license metadata. These indicators are combined into a readiness score ranging from 0 to 3. The score does not measure model quality or performance. It measures whether a model is accompanied by basic platform metadata that can support application-oriented reuse. We also use task family, reported size group, author type, and release cohort as descriptive or control variables when examining transformation likelihood.

The empirical analysis combines descriptive summaries, comparative statistics, and descriptive regression models. We summarize the distribution of public application transformation using linked-model shares, concentration curves, linked-Space counts, temporal lag distributions, and cohort-based transformation rates. We compare application transformation with visibility and technical reuse using Spearman rank correlations, top-k overlap, grouped application rates, and residual analysis. Specifically, an OLS model predicting log linked-Space counts from visibility and technical reuse indicators is used to identify models with more or fewer Space links than expected under conventional impact measures. To examine application readiness, we estimate logistic models predicting whether a model has at least one Model–Space link and report average marginal effects and adjusted predicted probabilities. These models are used for descriptive adjustment and comparison rather than causal identification.

## Results

### Platform structure, resource composition, and concentration

The Hugging Face ecosystem contains multiple repository types that have expanded together but at different scales. By the 2026 snapshot, the platform contained approximately 2.56 million model repositories, more than one million Spaces, and over 800,000 dataset repositories (Figure 1A). Structured Model–Space and Dataset–Space links also increased over the observation period, indicating that Spaces provide an observable layer in which models and datasets can be connected to public applications and demos. Yet this layer covers only a small fraction of the model population. Among all models in the snapshot, 57,414 models (2.24%), are linked to at least one Space (Figure 1B). Model release and Space linkage are therefore empirically distinct events: the expansion of model repositories does not translate proportionally into public application transformation. The Space layer also shows uneven resource composition. Most Spaces do not expose structured links to either models or datasets, while model-only Spaces account for a larger share than Spaces that combine both models and datasets (Figure 1C). Among the models that do enter this structured application pathway, linked-Space counts are highly concentrated. The top 10% of linked models account for 79.8% of all Model–Space links (Figure 1D). The full distribution is similarly skewed: most models have no linked Spaces, and the number of models declines sharply as linked-Space counts increase (Figure 1E). These patterns establish the basic empirical setting for the subsequent analyses: public application transformation is observable on Hugging Face, but it is sparse, selective, and highly concentrated.

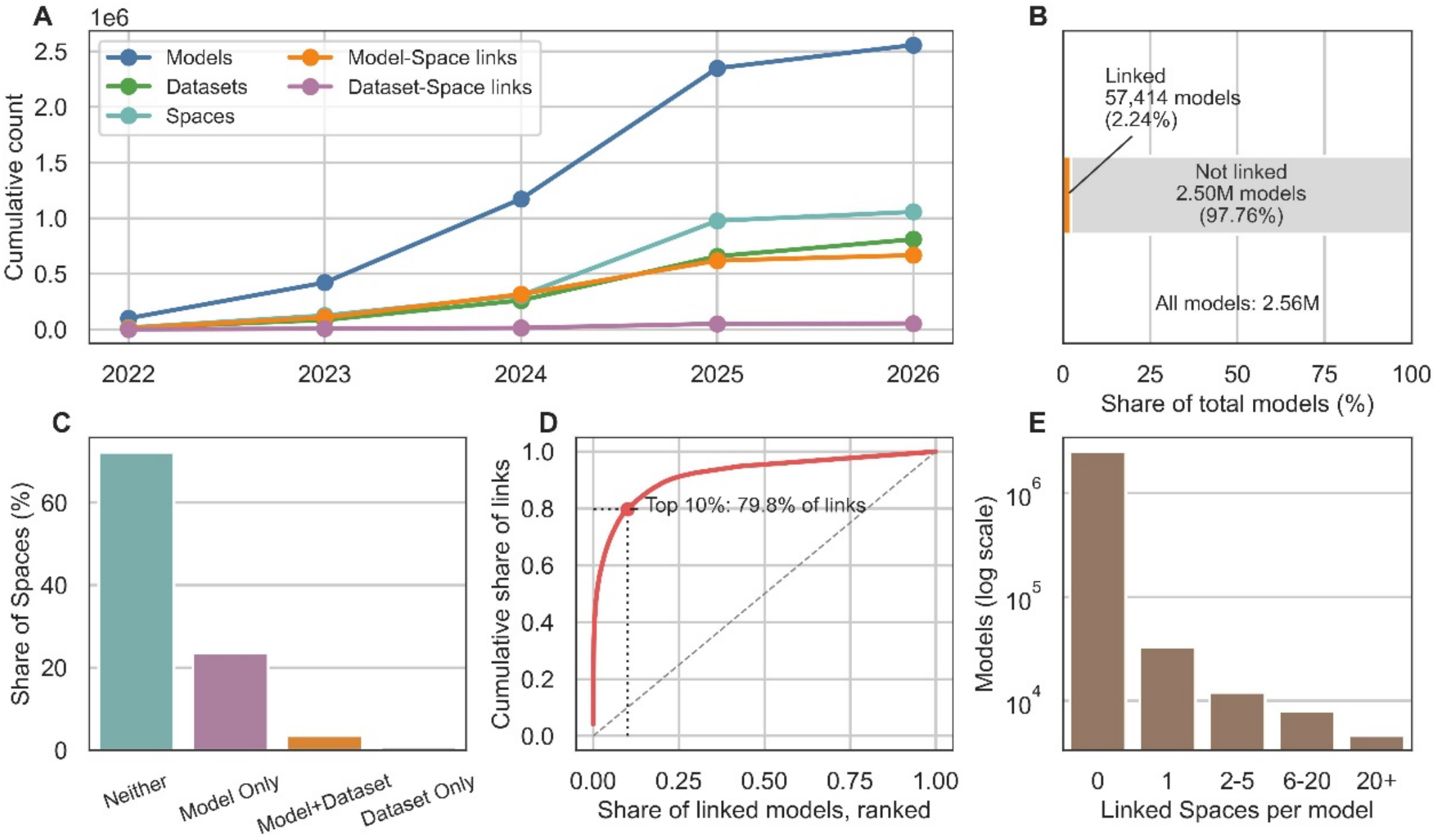

**Figure 1. Platform structure, resource composition, and concentration of public application transformation on Hugging Face. A.** Cumulative growth of models, datasets, Spaces, Model–Space links, and Dataset–Space links from 2022 to the 2026 snapshot. **B.** Share of all models linked to at least one Space, showing

that only a small minority of models enter the public application pathway. **C.** Resource composition of Spaces based on structured model and dataset links. **D.** Concentration curve of Model–Space links among linked models, with the top 10% of linked models accounting for 79.8% of all links. **E.** Distribution of linked Spaces per model on a log-scaled y-axis.

**Temporal dynamics of public application transformation**

Public application transformation expanded substantially during the main observation period. When Model–Space links are grouped by Space creation year, the number of observable Model–Space edges rises sharply from 2022 to 2025, alongside increases in both linked Spaces and linked models (Figure 2A). The decline in 2026 reflects the partial observation window rather than a full-year trend. Because the exact time when a model reference was added to a Space is not observed, these annual counts should be interpreted by Space creation year. Even with this constraint, the pattern shows that the Space layer became an increasingly active venue for connecting models to public applications and demos. The timing between model release and Space creation shows that application transformation is not limited to immediate post-release demonstrations. Across Space creation years, the distribution of model-to-Space lags shifts toward longer intervals, with later Spaces more often referencing models that are one to three years old or older (Figure 2B). The median lag and interquartile range also increase over time, indicating that Spaces increasingly draw on models released well before the Space itself was created (Figure 2C). This pattern points to delayed application use as an important feature of the public application pathway: models may enter Spaces shortly after release, but they may also remain available for later application-layer reuse.

The temporal structure of Model–Space links also varies by author relationship, task composition, and release cohort. Cross-author links account for the large majority of Model–Space edges throughout the observation period, rising from roughly four-fifths of links in 2022 to more than 90% in subsequent years (Figure 2D). This indicates that Spaces are not only self-hosted demonstrations by model publishers, but also frequently involve models incorporated by other accounts. The task composition of Model–Space links shifts over time as well, with text-generation and text-to-image models becoming increasingly prominent among linked models (Figure 2E). Fixed-window cohort transformation rates further show that models released in earlier cohorts have higher observed probabilities of entering Spaces within 30, 90, 180, and 365 days than more recent cohorts (Figure 2F). Taken together, the temporal results show that public application transformation is both uneven and delayed: it grows over time, often involves cross-author use, and does not operate only as an immediate response to newly released models.

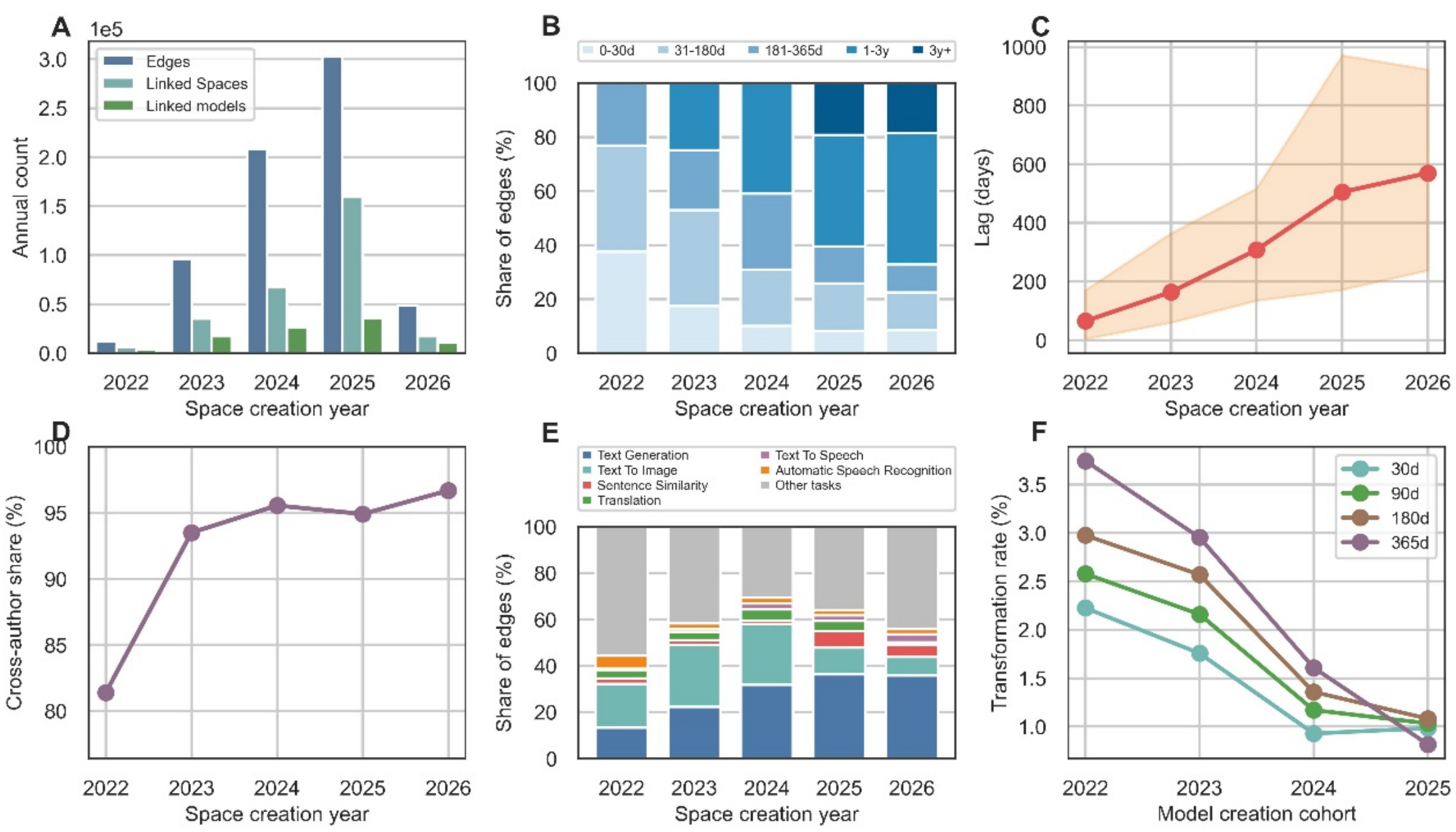

**Figure 2. Temporal dynamics of public application transformation through Hugging Face Spaces. A.** Annual counts of Model–Space edges, linked Spaces, and linked models by Space creation year. **B.** Distribution of model-to-Space lag categories by Space creation year, measured as the difference between model creation date and Space creation date. **C.** Median lag and interquartile range of model-to-Space timing by Space creation year. **D.** Share of Model–Space edges that connect models and Spaces created by different authors. **E.** Task composition of Model–Space edges by Space creation year. **F.** Model cohort transformation rates within fixed post-release windows of 30, 90, 180, and 365 days.

**Application transformation as a distinct impact dimension**

We next examine whether public application transformation reproduces existing indicators of model impact. The rank-correlation results show that Model–Space links are related to platform visibility and technical reuse, but only moderately so (Figure 3A). Space links correlate positively with downloads, likes, community visibility, and technical reuse, indicating that application-transformed models are not disconnected from broader platform attention or model-production reuse. The association is stronger with likes than with technical reuse, while the correlation between total Space links and cross-author Space links is much higher, suggesting that the Model–Space signal is not driven only by self-authored demonstrations. The top-k overlap analysis reinforces this pattern (Figure 3B). Among the top application-transformation models, overlap is higher with high-community visibility models than with technically reused models, but the overlap remains incomplete across the top 1%, 5%, and 10% thresholds. These results indicate that Model–Space links capture a related but non-identical form of model impact.

The joint distribution of community visibility and technical reuse further clarifies how these dimensions combine. In Figure 3C, visibility groups are defined by quartiles of the community visibility score: models in the bottom quartile are coded as low visibility, models in the second and third quartiles are grouped as middle visibility, and models in the top quartile are coded as high visibility. Technical reuse is grouped separately because the distribution is highly zero-inflated. Models with no technical reuse form a distinct no-reuse group, while models with positive technical reuse are divided into quartile-based groups: low positive reuse (Q1), middle positive reuse (Q2–Q3), and high positive reuse (Q4). Application rates are highest when high community visibility is combined with high positive technical reuse. Within the high-visibility group, models with no technical reuse have an application rate of 6.31%, while models in the high positive-reuse group reach 80.76%. At the same time, high community visibility alone does not guarantee application transformation, and lower-visibility groups remain much less likely to enter Spaces. This pattern shows that application transformation is associated with both attention and model-production reuse, but remains selective within these categories.

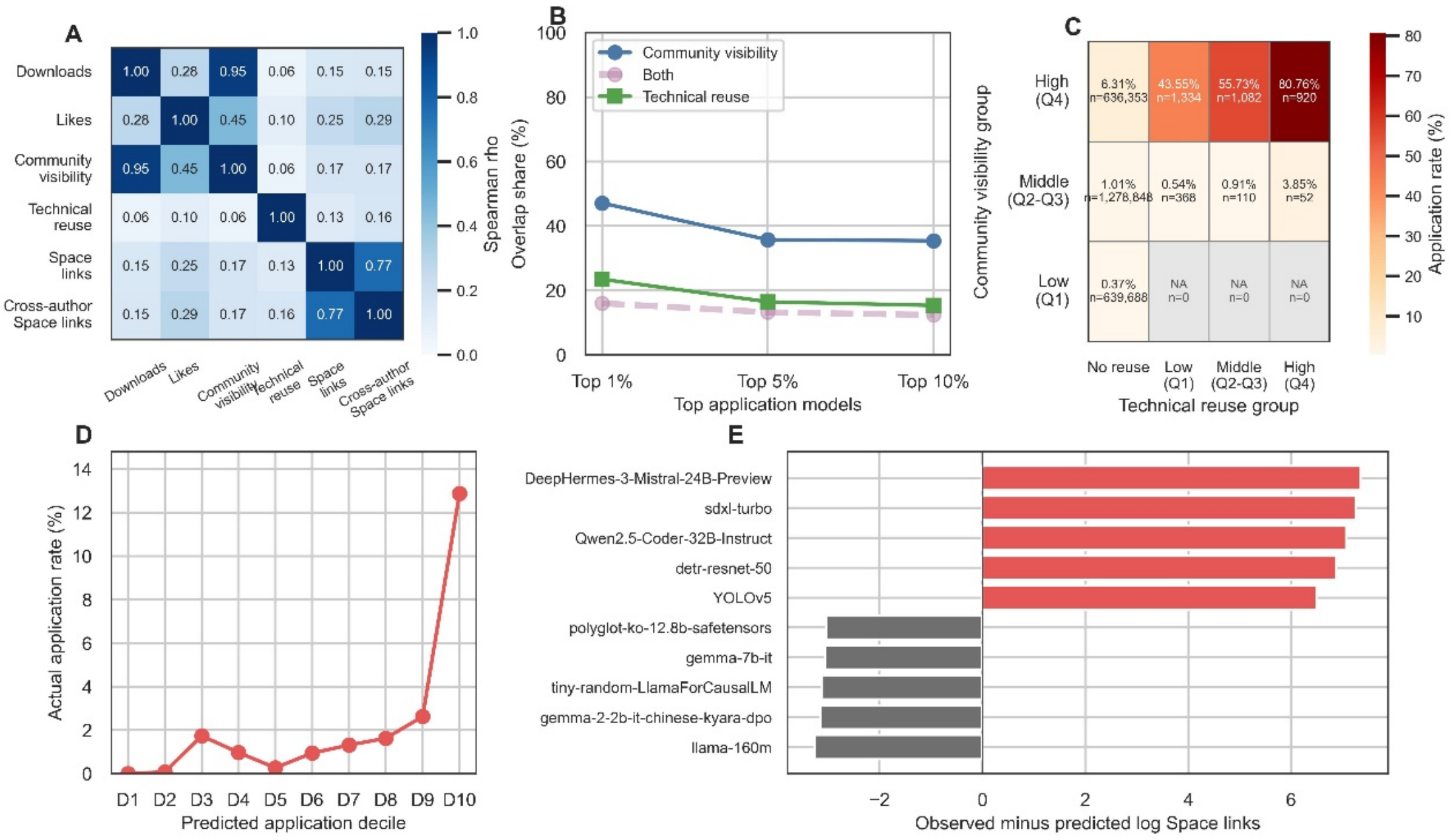

**Figure 3. Public application transformation compared with visibility and technical reuse. A.** Spearman rank correlations among downloads, likes, community visibility, technical reuse, total Space links, and cross-author Space links. **B.** Overlap between top application-transformation models and top models by visibility, technical reuse, or both. **C.** Application rate across community visibility groups and technical reuse groups, showing that high visibility and positive technical reuse jointly correspond to higher application transformation. **D.** Actual application rate by deciles of OLS-predicted application transformation, based on downloads, likes, and technical reuse. **E.** Models with the largest positive and negative residuals from the OLS prediction of log Space links.

To examine whether conventional impact indicators fully account for application transformation, we use a descriptive residual analysis. We first estimate an OLS model predicting log linked-Space counts from downloads, likes, and technical reuse, and then group models into deciles based on their predicted values. The actual application rate rises sharply in the highest predicted decile, while most lower deciles remain much lower (Figure 3D). This indicates that visibility and technical reuse explain part of the variation in Model–Space linkage. However, the residual plot shows that substantial deviations remain (Figure 3E). Positive residuals identify models that receive more Space links than predicted by their downloads, likes, and technical reuse, including models such as DeepHermes-3-Mistral-24B-Preview, SDXL-Turbo, Qwen2.5-Coder-32B-Instruct, DETR-ResNet-50, and YOLOv5. Negative residuals identify models with fewer Space links than predicted, including Gemma-7B-it, tiny-random-LlamaForCausalLM, Gemma-2-2B-it-Chinese-Kyara-DPO, and Llama-160M. These residual cases are not interpreted causally. They show that application transformation contains model-level variation not fully captured by conventional impact indicators.

Taken together, these comparisons show that public application transformation is neither a random platform artifact nor a simple restatement of conventional impact indicators. It is associated with community visibility and technical reuse, as shown by the grouped application rates and predicted deciles, but the incomplete top-k overlap and residual cases indicate that these indicators do not fully account for which models enter Spaces. Public application transformation is therefore visibility-dependent, but it is not reuse-equivalent. It captures an application-facing profile of model impact that is not captured by downloads, likes, or model-to-model technical reuse alone.

**Application readiness and transformation likelihood**

Having shown that public application transformation is not reducible to community visibility or technical reuse, we next examine the model-level conditions associated with entering Spaces. We begin with metadata-based readiness. Readiness is measured through three basic platform metadata indicators: whether a model reports task metadata, library metadata, and license metadata. Models linked to at least one Space have substantially higher coverage on all three indicators than models without Model–Space links (Figure 4A). The difference is especially visible for task and license metadata, indicating that application-transformed models tend to be more clearly described and more legible as reusable platform objects. We combine these three indicators into a readiness score ranging from 0 to 3, which is used below to examine whether more complete metadata is associated with higher transformation likelihood.

Application transformation varies across task families and reported model size. In Figure 4B, model tasks are grouped into broad task families, and reported-size groups are defined within task families, with models that do not report size kept as a separate category. The heatmap shows clear heterogeneity across tasks. Generative vision models have relatively high application rates, especially among small reported-size models, while text-generation and broader Text/NLP models show lower rates across size groups. Task-level application rates provide a complementary view (Figure 4D). Among tasks retained for task-level comparison, image-to-image, translation, zero-shot image classification, text-to-speech, and text-to-video show application rates well above the overall baseline. These patterns are consistent with the idea that task affordances matter for public application transformation. Models whose outputs are easier to demonstrate, compare, or interact with in a Space appear more likely to enter the application layer. Author type further conditions application transformation, but mainly among highly visible models. In the raw comparison, organization-owned models have higher application rates than individually owned models in the high-visibility group, while the difference is much smaller in low- and middle-visibility groups (Figure 4C). This pattern indicates that organizational authorship is not associated with a uniform advantage across the platform. Rather, the organization–individual difference becomes most visible among models that already receive substantial platform attention.

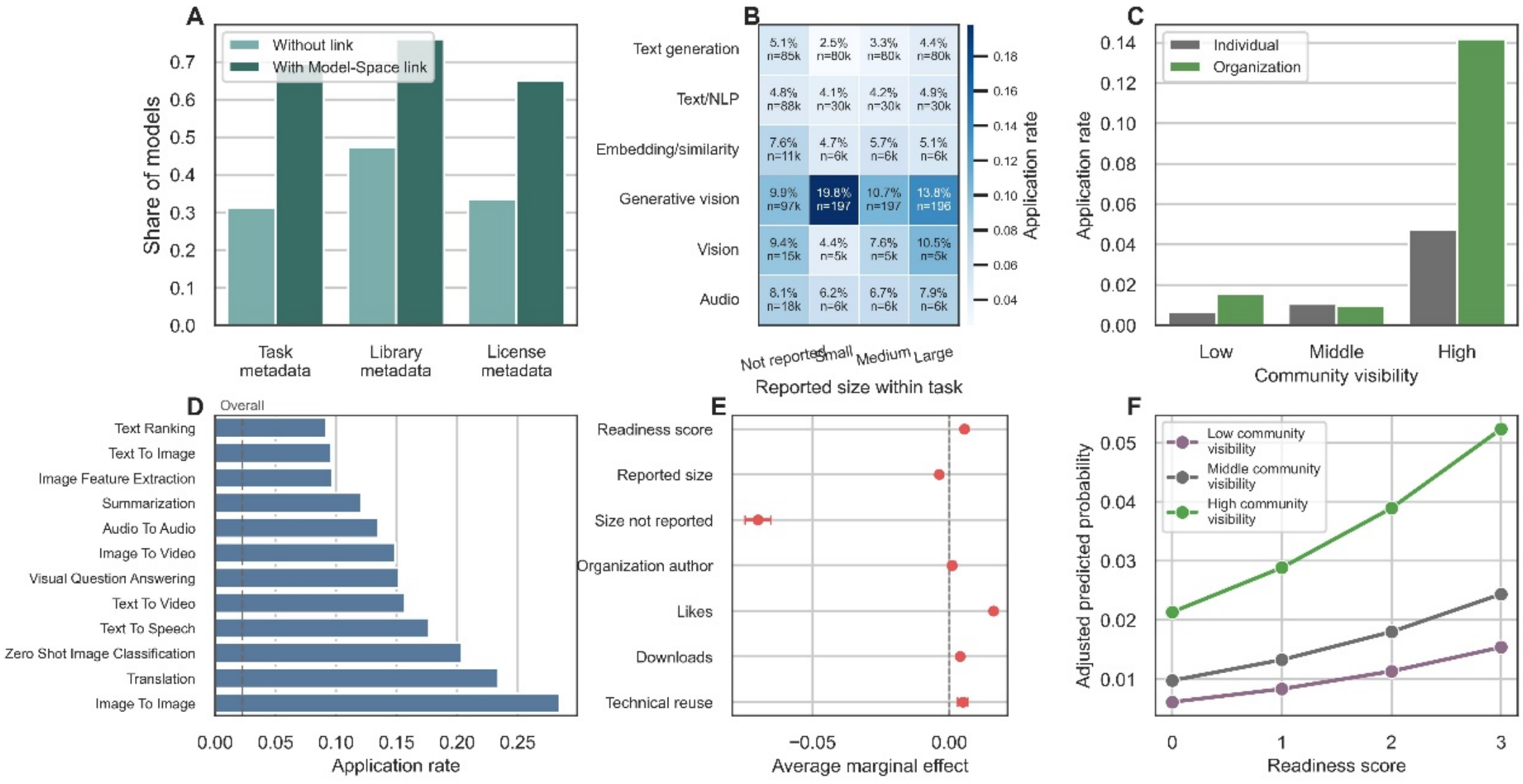

**Figure 4. Application readiness and likelihood of public application transformation. A.** Coverage of task, library, and license metadata among models with and without Model–Space links. **B.** Application rates across task families and reported model-size groups. **C.** Application rates by community visibility group and author type. **D.** Task-level application rates, with the dashed line indicating the overall application rate. **E.** Average marginal effects from a logistic model predicting whether a model is linked to at least one Space. **F.** Adjusted predicted probability of application transformation by readiness score across visibility groups.

Because readiness, visibility, technical reuse, size reporting, and authorship are correlated at the model level, we use a logistic model to examine their adjusted associations with application transformation. The model predicts whether a model has at least one Model–Space link and reports average marginal effects for readiness score, reported size, an indicator for missing size information, organization authorship, likes, downloads, and technical reuse (Figure 4E). Readiness score remains positively associated with application transformation after adjustment, as do likes, downloads, and technical reuse. Models without reported size information are less likely to enter Spaces, suggesting that missing size metadata is associated with lower application-layer uptake. By contrast, the adjusted association for organization authorship is comparatively limited once community visibility, reuse, readiness, size, and cohort differences are included. The adjusted prediction plot translates this pattern into transformation probabilities across visibility groups (Figure 4F). Within each visibility group, predicted transformation probability increases with readiness score, and the increase is largest among high-visibility models. At the same time, the predicted probabilities remain modest even for high-visibility models with full readiness metadata. Taken together, these results indicate that public application transformation is associated with both informational readiness and platform attention, but transformation remains selective even among visible and well-documented models.

**Application pathways and resource assemblages**

Public application transformation does not end when a model becomes linked to a Space. Once models enter the Space layer, they are placed into specific application configurations that may involve datasets, hosting environments, and task-specific application categories. At the Space level, most model-linked Spaces are model-only configurations, while a smaller share combine models with one or more datasets or include multiple models together with dataset resources (Figure 5A). This pattern indicates that public application transformation is primarily model-centered, but not limited to isolated model demonstrations. A subset of Spaces functions as a broader resource assemblage in which models are combined with datasets and application environments. After models become linked to public applications or demos, they can therefore appear either as standalone components or as part of multi-resource configurations.

The edge-level analysis shows that different model tasks enter different metadata-derived application categories. In Figure 5B, application categories are derived from Space tags and related metadata, and the heatmap reports the distribution of Model–Space edges across these categories within each model task family. Text-generation models are distributed mainly across agent/tool and data/benchmark categories, while text-to-image models are more concentrated in agent/tool-oriented Spaces. Sentence-similarity models are strongly associated with data/benchmark Spaces, and translation models are primarily linked to Text/NLP applications. The Sankey-style flow provides a volume-sensitive summary of the same task-to-category relationship (Figure 5E). It shows that the largest flows come from text-generation and other-task models into agent/tool and data/benchmark categories, while more specialized tasks such as translation, sentence similarity, text-to-speech, and automatic speech recognition follow narrower application pathways. These patterns suggest that application transformation is organized through task-specific pathways rather than a uniform model-to-Space route.

Dataset co-use and hosting environments further clarify how these application pathways are configured. At the Space level, dataset co-use is defined as whether a model-linked Space also contains at least one structured Dataset–Space link. Dataset co-use varies substantially across metadata-derived application categories. It is highest in agent/tool and data/benchmark Spaces, and much lower in image/vision, chatbot/conversation, audio, and Text/NLP categories (Figure 5C). Hosting environments are also unevenly distributed. Gradio accounts for the large majority of model-linked Spaces, while Docker and Streamlit form smaller but visible hosting pathways, and Static or other/missing SDK categories account for only a limited share (Figure 5D). Taken together, these results shift the interpretation of public application transformation from a binary model-level outcome to a set of heterogeneous Space-level and edge-level configurations. Models do not simply enter Spaces; they are organized into different application categories, dataset-use patterns, and hosting environments.

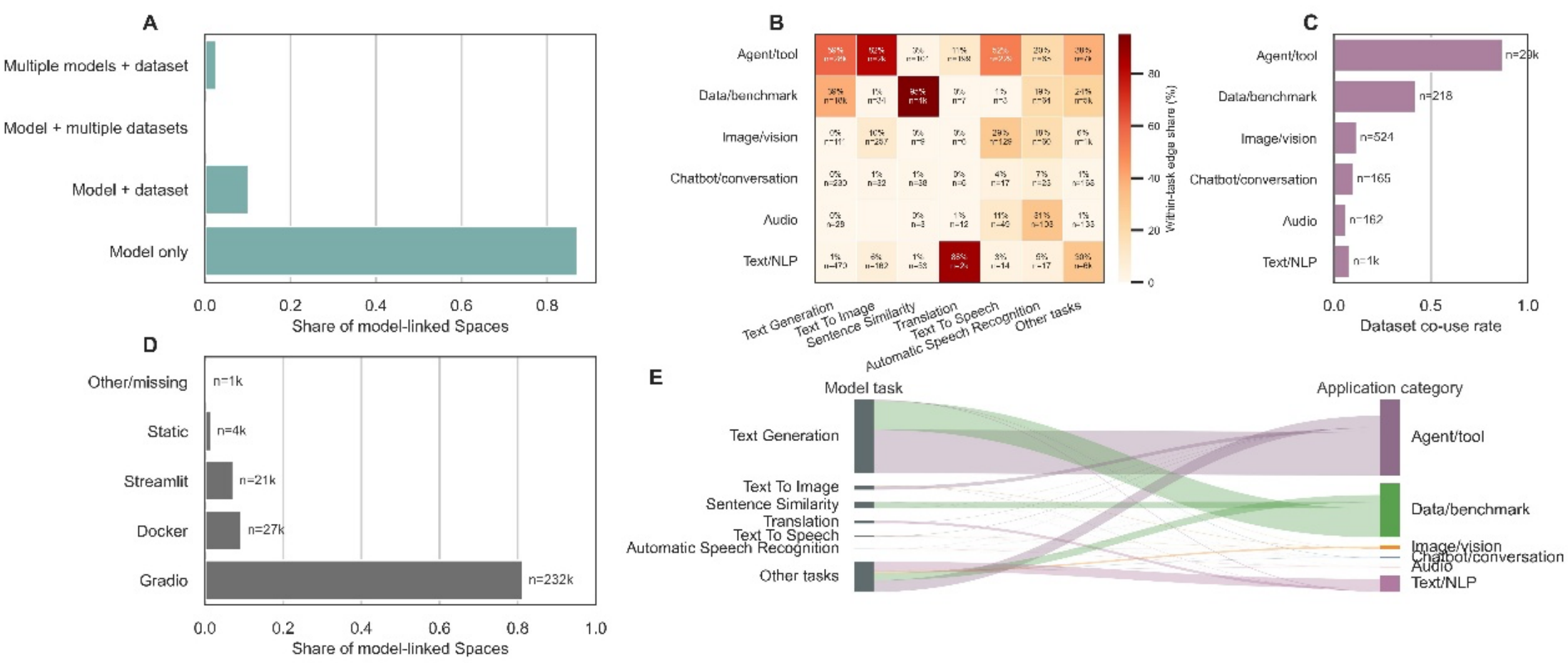

**Figure 5. Application pathways and resource assemblages after public application transformation. A.** Resource composition of model-linked Spaces, distinguishing model-only Spaces from configurations that combine models with one or more datasets. **B.** Distribution of Model–Space edges across metadata-derived application categories within each model task family. **C.** Dataset co-use rate by application category among model-linked Spaces. **D.** SDK distribution of model-linked Spaces, showing the hosting environments used to build public applications and demos. **E.** Sankey-style flow from model task families to metadata-derived application categories, showing differentiated application pathways after models enter Spaces.

**Discussion**

In this study, we examined how open-source models move from repository release to public applications and demos in the Hugging Face ecosystem. Using Model–Space links as platform-visible traces, we measured public application transformation and compared it with community visibility, technical reuse, application readiness, and resource assemblage. The results show that this transformation is sparse and concentrated, related to community visibility but not reducible to technical reuse, and associated with metadata-based readiness signals. They also show that models entering Spaces are not organized through a single pathway, but through heterogeneous configurations involving datasets, SDKs, author relationships, and task-specific application categories. These findings shift attention from whether models are merely released or popular to how they travel across platform layers after release.

Model release should therefore be understood as the beginning of a platform trajectory rather than as evidence of impact. Prior studies have shown that Hugging Face has become a major infrastructure for open-source model development, with models, datasets, Spaces, user activity, and documentation practices forming a rapidly expanding ecosystem (Osborne et al., 2024; Sangari et al., 2025). Our findings show that this expansion does not translate automatically into application-layer transformation. A model can be released and remain visible as a repository object, but entering a public Space marks a different platform transition: the model becomes part of an interface, demo, tool, benchmark app, or resource configuration that can be observed beyond the repository itself.

The main measurement implication is that open-source model impact should be treated as multidimensional. The broader platform-metrics literature has long cautioned that visibility, popularity, adoption, and reuse indicators capture different forms of platform activity rather than a single underlying hierarchy (Saini et al., 2020; Zerouali et al., 2019). Existing model reuse studies have also made important progress in tracing how models are fine-tuned, adapted, referenced, or incorporated into downstream code and software repositories (Jiang et al., 2024; Palombo et al., 2025; Taraghi et al., 2024). Model–Space links complement these indicators by identifying a different layer of circulation. Downloads and likes capture platform attention, and model-to-model relations capture reuse in the model production layer. Public application transformation captures whether models become incorporated into public application-facing artifacts. It is therefore visibility-dependent, but not reuse-equivalent.

The readiness results shift attention from model capability alone to the informational conditions that make models movable across platform layers. Model cards and related documentation have been widely discussed as mechanisms for transparency, accountability, and responsible reporting (Mitchell et al., 2019). Studies of model repositories further show that intended-use information, metadata, and repository descriptions are often incomplete or unevenly reported (Gong et al., 2023; Li, Kant, et al., 2023). The present findings suggest an additional role for such information. Task, library, and license metadata are not only descriptive features of repositories; they are also part of the readiness conditions associated with application transformation. The results do not show that metadata causes models to enter Spaces, but they indicate that application-transformed models tend to be more legible, integrable, and clearly governed as platform objects.

At the application layer, Spaces make visible the assembled character of public AI demos and applications. System and demo papers illustrate how Hugging Face Spaces can host interactive model- or data-oriented interfaces (Arora et al., 2025). The ecosystem-level results show that such Spaces are not uniform endpoints. Some are model-only demonstrations, while others combine models with datasets, SDKs, author relationships, and task-specific application categories. Public application transformation is therefore not simply a binary outcome in which a model either enters or does not enter a Space. After entry, models are organized into different resource assemblages, and their application-layer impact depends partly on how they are packaged, linked, hosted, and combined with other platform resources.

Several limitations qualify these findings. First, Spaces represent public applications and demos hosted on Hugging Face, not all forms of downstream deployment or adoption. Model–Space links do not capture private deployments, API-based uses, or applications hosted outside the platform. Second, the analysis depends on structured metadata and may miss code-level dependencies, external API calls, manually embedded model usage, and other unstructured forms of reuse. Timing is also approximate because Space creation date does not necessarily equal the time when a model reference was added. Third, the regression analyses are descriptive rather than causal. Associations among readiness, visibility, technical reuse, and application transformation may reflect unobserved model quality, developer selection, platform attention, or metadata completion after uptake. Future research could extend this work by combining platform metadata with code-level dependency extraction, finer-grained temporal traces, external deployment records, or qualitative studies of how developers select and assemble models into public applications.